\documentclass[runningheads]{llncs}

\usepackage[T1]{fontenc}
\usepackage{xcolor,color}
\usepackage{graphicx}
\usepackage{amsmath,amssymb,mathrsfs}
\usepackage{tikz}
\usetikzlibrary{arrows.meta,decorations.pathreplacing,calc}
\usepackage{booktabs}
\usepackage{subcaption}
\usepackage{placeins}
\usepackage{colortbl}
\usepackage{wrapfig}
\usepackage{enumitem}
\newlength{\parheadskip}
\newcommand{\parhead}[1]{\vspace{\parheadskip}\noindent\textbf{#1}}

\usepackage[hidelinks]{hyperref}
\begin{document}

\title{Active Liquidity On Chain:\\ Evidence from PropAMMs Across Chains}
\titlerunning{Active Liquidity On Chain: Evidence from PropAMMs Across Chains}
\author{Ozan Solmaz\inst{1} \and Lioba Heimbach\inst{2} \and Jason Milionis\inst{2}}
\authorrunning{O. Solmaz et al.}
\institute{ETH Zurich \and Category Labs}
\maketitle

\begin{abstract}
The liquidity providers of most typical automated market makers (AMMs) are passive and known to suffer from adverse selection. AMMs traditionally rely on trades executed on them to sync the price with external markets (and a notion of fair value thereof). As a result, when an external fair value moves, arbitrageurs trade on AMMs to pick off their stale quotes. A recent approach termed \textit{proprietary automated market makers (propAMMs)} emerged as a response in 2024: these on-chain programs instead have their singular operator quote from its own inventory, repricing without a trade via efficiently provided price updates. By 2026, propAMMs accounted for more than half of SOL/USDC volume on Solana.

We present the first year-long longitudinal measurement of propAMMs on Solana, Base and Monad, and decode the on-chain logic of the dominant propAMM on Base. We find that two seconds after a fill, propAMMs earn $0.37$\,bps on Solana and $1.19$\,bps on Base, while AMMs lose $0.22$ and $0.62$\,bps. We empirically quantify and classify their edge as stemming from four factors: propAMMs continuously reprice rather than waiting for a trade, they charge for the source-dependent risk each counterparty might bring, they avoid cross-venue arbitrage from other on-chain markets (such as other AMMs), and they spoof by filling trades at a worse price than they quote. Finally, we show that on our proxy for retail flow (i.e., fills that arrive when the reference price is not moving) propAMMs collect $0.26$\,bps on Solana where AMMs collect $2.59$\,bps, indicating that they offer better prices for short-term uninformed flow.

\end{abstract}

\section{Introduction}
\textit{Decentralized finance (DeFi)} emerged on Ethereum during the DeFi summer of 2020, at a time when on-chain transaction costs were significant and throughput was limited. During this time, the type of \textit{decentralized exchange (DEX)} that took off was the \textit{automated market maker (AMM)}, which importantly keeps on-chain transaction and storage costs low, as the price it offers generally follows from an invariant over the pooled reserves. Thus, unlike in a traditional limit order book, there are no individual orders to store, match or cancel on chain.

Liquidity providers on most AMMs are passive and their offered price quotes in response to requested trades only update as a result of a trade (as opposed to information coming externally to trades). When an asset's fair value (such as---but not only---an external market price) moves, arbitrageurs therefore trade against the stale quotes on AMMs, i.e., liquidity providers are adversely selected. As a result, they suffer an economic loss known as \textit{loss-versus-rebalancing (LVR)}, which reflects the staleness of an AMM's quote relative to a fair value~\cite{milionis2026automatedmarketmakinglossversusrebalancing}.

In 2024, \emph{proprietary automated market makers (propAMMs)} started to emerge on Solana. A propAMM is an on-chain program whose pricing function executes in a smart contract, as in an AMM, but whose prices are set by a single operator. The operator supplies the inventory themselves and (importantly) revises the quoted prices by writing new parameters to the contract, i.e., without a trade. As a result, propAMMs belong to paradigms offering active liquidity management, more akin to how a limit order book operates, despite being a hybrid. PropAMMs thereby attempt to keep on-chain costs low while avoiding adverse selection. Active quoting also brings about tighter spreads and thereby better prices to traders. By 2026, propAMMs accounted for more than half of SOL/USDC volume on Solana~\cite{lostin2025propamm}.

We perform the first longitudinal measurement study of propAMMs across three chains, i.e., Solana, Base and Monad. We measure the extent to which propAMMs have an edge over liquidity providers on AMMs, where that edge comes from, and its effect on retail flow.

\parhead{Our contributions.} We summarize our main contributions as follows: 
\begin{itemize}[topsep=0pt]
    \item To the best of our knowledge, we are the first to perform a longitudinal multi-chain measurement study of the propAMM landscape, showing that propAMMs account for a large share of volume of decentralized trading on Solana, Base and Monad.
    \item We measure markouts and find that unlike passive liquidity providers on AMMs, most propAMMs have positive markouts and thus successfully and selectively avoid arbitrageurs attempting to pick off their quotes. Two seconds after a fill, propAMMs earn $0.37$\,bps on Solana and $1.19$\,bps on Base, while AMMs lose $0.22$ and $0.62$\,bps.
    \item We decode the closed-source pricing logic of three propAMMs on Base, i.e., Tessera in full and Metric and Elfomo in part, to show how propAMMs price.
    \item Combining our measurements with the decoded contract logic, we trace where the edge that propAMMs have over passive liquidity providers comes from, and we highlight four sources: (1) frequent and inexpensive quote updates, (2) pricing by counterparty, i.e., charging non-retail flow more than aggregator flow, (3) being on the winning leg of cross-venue arbitrage (such as when a propAMM and an AMM are jointly arbitraged against each other), and (4) spoofing, i.e., quoting better prices than a trade actually receives.
    \item We approximate retail flow by fills that arrive when the reference price is not moving. On this flow propAMMs collect $0.26$\,bps on Solana, $1.26$\,bps on Base and $1.62$\,bps on Monad, against $2.59$, $1.38$ and $8.60$\,bps on AMMs, and retail makes up a larger share of propAMM notional on all three chains.
\end{itemize}

\section{Background}
\subsection{Automated Market Makers}
An \textit{automated market maker (AMM)} generally prices trades so that a fixed function of the pool's token reserves remains constant, after accounting for trading fees~\cite{adams2020uniswapv2,adams2021uniswapv3,angeris2023geometryconstantfunctionmarket}. By avoiding the need to store and maintain individual limit orders, AMMs can reduce the storage and computation required for on-chain exchange. This efficiency helped motivate their use in the early days of DeFi~\cite{milionis2024complexity,angeris2022multiasset}.

Anyone can become a liquidity provider by depositing the pool's tokens. Traders swap against the pool's reserves and pay a fee on their input, which accrues pro rata to the pool's liquidity providers and, depending on the protocol, to the protocol~\cite{adams2025unification}. The most common invariant is the constant product used by Uniswap, applied either across the full price range or, with concentrated liquidity, within a price range chosen by each liquidity provider. Prices are thus a function of the reserves alone and move only once a trade has occurred. AMMs thus rely on arbitrage trades to align pool prices with external markets,\footnote{LPs in a Uniswap v2-style pool can also change its marginal price without a swap by withdrawing liquidity and redepositing assets at a different reserve ratio.} exposing liquidity providers to adverse-selection costs associated with LVR~\cite{milionis2026automatedmarketmakinglossversusrebalancing,nezlobin2025lossversusrebalancingdeterministicgeneralizedblocktimes,wu2025measuringcexdexextractedvalue}.

\subsection{Proprietary Automated Market Makers}
In a propAMM, the operator reprices liquidity directly through parameter updates, i.e., without a swap. The pricing logic is proprietary, as the name indicates. We nonetheless were able to decode the on-chain logic of Tessera on Base, the largest propAMM on Base by volume and among the three largest on Solana, as well as the logic of the Metric and Elfomo implementations used in August 2026. Next, we describe the operation of a propAMM based on Tessera's Base implementation and detail the differences to Metric and Elfomo in Appendix~\ref{app:base_prop_amm}. Note that this description serves as an illustrative example of how the on-chain logic of propAMMs operates, and that implementations we do not decode may differ.

\begin{figure}[t]\vspace{-10pt}
    \centering
    \includegraphics[width=\linewidth]{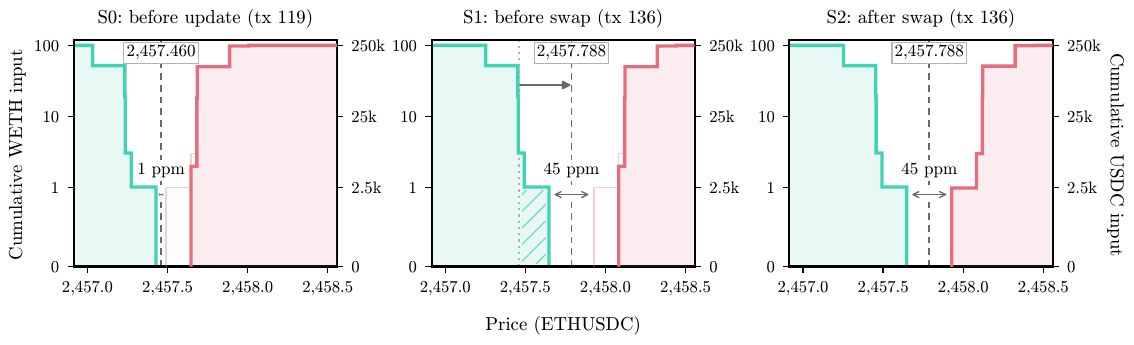}\vspace{-8pt}
\caption{Tessera's WETH/USDC order book evolution in block 50,369,777 on Base. Green is the bid ladder, red the ask ladder. Faint curves give configured liquidity, bold curves the accessible liquidity after netting. First, an update transaction at index \href{https://basescan.org/tx/0xddd0d399d1b606bccf40fa570294cbe6e55f296b5cc40953091a46cb1d6fff70}{119} raises the anchor price $p$ from 2,457.460 to 2,457.788 and widens the spread $s$ from 1 to 45~ppm ($S_0$ to $S_1$). Then, a swap transaction at index \href{https://basescan.org/tx/0x4230f2dbc40c4c202064e6d359f2978a105ca33866c43e558582aa3768a3013c}{136} sells 1.0177~WETH. In $S_1$ the hatched area marks the input of the swap. As the incoming WETH is netted against earlier USDC flow, the bid side does not change, while earlier ask levels become available again in $S_2$.}\vspace{-10pt}
    \label{fig:tess_exec}
\end{figure}

Tessera represents liquidity through two directional ladders with level prices determined by an \textit{anchor price}, a \textit{spread}, and level-specific price multipliers.\footnote{Our mechanism description refers to implementation \href{https://basescan.org/address/0x6d9DD143e42b6338F4F6a7c0C26D124658F641CB}{\texttt{0x6d9...41CB}}, activated in the inspected pools on 12 Aug 2026, which introduced the dedicated \textit{spread} parameter.} Panel $S_2$ of Figure~\ref{fig:tess_exec} shows such a book, with the anchor price marked by the dashed line and the spread as the gap between the two ladders. The bidirectional ladders resemble the two sides of an order book. In a traditional order book, each level stores its own price. When a market maker's belief about the true price moves, the market maker must thus cancel and replace every order. Tessera instead stores a single anchor price $p$ and a single spread $s$, and gives each level a fixed multiplier that places it relative to the two. Adjusting $p$ and $s$ reprices every level at once ($S_0$ to $S_1$ in Figure~\ref{fig:tess_exec}), which reduces storage access count and thus saves on-chain costs. Individual levels can also be adjusted by changing their multipliers. This is relatively costly in comparison (26,706 median gas for an update of $p$ and $s$ on Tessera vs.\ 152,013 median gas for updating the book levels) and is therefore rare compared to updates of $p$ and $s$ in our data sample (1 price update per 0.563 blocks vs.\ 1 book update per 4.31 blocks). The propAMM can additionally exclude a configurable number of levels closest to the anchor price, which withdraws the best prices without replacing the book.

Liquidity accounting also avoids per-order book level writes. The propAMM keeps two running totals, one per asset, of the amounts traded in each direction. From this it derives the remaining liquidity. Trading in one direction can deplete the levels closest to the anchor price or replenish the levels in the opposite direction ($S_1$ to $S_2$ in Figure~\ref{fig:tess_exec}). Thus, if a trade depleted the closest level, later trades in the same direction reach deeper into the order book ladder and receive worse prices. The two directions are netted at the current anchor price $p$.

A trade swapping $x$ tokens of asset $X$ executes as follows. It first walks the available levels of the order book, which yields the output the book alone would give ($Y_{\mathrm{book}}(x)$). The propAMM then withholds a share $\phi$ of that output. Thus, the swap returns $Y_{\mathrm{exec}}(x) = Y_{\mathrm{book}}(x)(1-\phi)$. Here, the penalty $\phi$ depends on, among other things, how old the quote is, where the order flow is coming from (e.g., a whitelisted aggregator vs. potentially toxic flow) and the priority fee of the transaction. This hints at where the propAMM edge  comes from (Section~\ref{sec:edge}).

\section{Related Work}
A large literature studies liquidity provider profitability in DeFi. The first such measure was ``impermanent loss,'' i.e., the difference in value between a liquidity provider's position and simply holding the deposited assets. Measurements for Uniswap v2- and v3-style pools show that liquidity providers generally lose in this sense~\cite{heimbach2021behavior,loesch2021impermanent,heimbach2022risks}. Fan et al.~\cite{fan2023strategic} and Fritsch~\cite{fritsch2021concentrated}, on the other hand, study how to provide liquidity strategically in order to reduce such losses.

Formally, \textit{loss-versus-rebalancing (LVR)} isolates the true economic cost arising precisely from the fundamental price staleness of AMMs and the corresponding arbitrage, as introduced by Milionis et al.~\cite{milionis2026automatedmarketmakinglossversusrebalancing}. Subsequent work studied the role of trading fees and finite block times~\cite{milionis2023automated}, and extended their framework to general block time distributions~\cite{nezlobin2025lossversusrebalancingdeterministicgeneralizedblocktimes}. Fritsch and Canidio~\cite{fritsch2024measuring} measured the loss empirically, validating theoretical predictions of Milionis et al.~\cite{milionis2023automated}, and found that it frequently exceeds the fees liquidity providers collect. A framework for active liquidity effects on AMMs was also given by Milionis et al.~\cite{milionis2023flairmetricliquidityprovider}.

A parallel line of work proposes AMM designs with the aim of reducing LVR. Diamond~\cite{mcmenamin2022diamonds} auctions the arbitrage opportunity to the block producer, the am-AMM~\cite{adams2024amamm} auctions the right to manage the pool, and Canidio and Fritsch~\cite{canidio2023arbitrageurs} batch orders at a single clearing price. These designs keep liquidity provision passive. In contrast, we  study propAMMs that rely on active liquidity provision and quote updating to avoid adverse selection. In this sense, they are somewhat more similar to limit order books than AMMs, despite being a hybrid of both.

Heimbach et al.~\cite{heimbach2024nonatomic} first measure CEX--DEX arbitrage on Ethereum. Wu et al.~\cite{wu2025measuringcexdexextractedvalue} quantify the value these arbitrageurs extract from the liquidity providers of AMMs by marking out their fills against CEX prices. We measure markouts from the perspective of propAMMs, and show that they have positive markouts.

\section{Data Collection}

We study propAMMs on Solana, Base and Monad. PropAMMs first appeared on Solana, where they remain largest, and have since launched on Base and Monad.\footnote{We exclude Ethereum, as propAMMs there only launched in May 2026 and rely on block builders ordering quote updates ahead of trades~\cite{fuller2026propamms}. Thus, they behave more like RFQs than propAMMs.}
Our data spans the year from 1 September 2025 to 31 August 2026, or starts later where propAMMs launched on a chain after that date. We collect propAMM trades and updates, along with AMM swaps on Solana, Base and Monad. Additionally, we collect Bybit order book data for comparison. Appendix~\ref{app:data_collection} provides a detailed description of our pipeline.

\begin{figure}[t]\centering\vspace{-12pt}
\begin{subfigure}[b]{2.0in}\centering
  \includegraphics[scale=1]{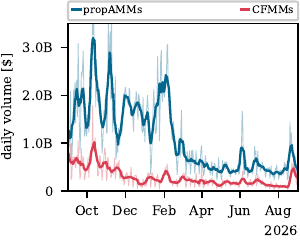}\vspace{-8pt}
  \caption{Solana SOL/USDC: volume}
\end{subfigure}\hspace{10pt}%
\begin{subfigure}[b]{2.34in}\centering
  \includegraphics[scale=1]{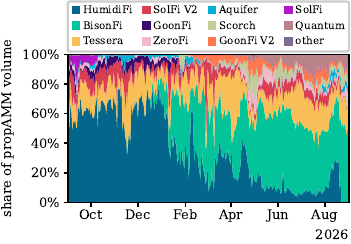}\vspace{-8pt}
  \caption{Solana: share by propAMM}
\end{subfigure}\\[4pt]
\begin{subfigure}[b]{2.0in}\centering
  \includegraphics[scale=1]{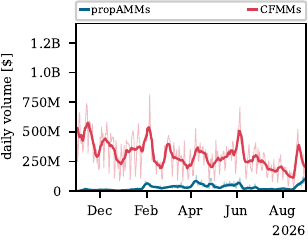}\vspace{-8pt}
  \caption{Base ETH/USDC: volume}
\end{subfigure}\hspace{10pt}%
\begin{subfigure}[b]{2.34in}\centering
  \includegraphics[scale=1]{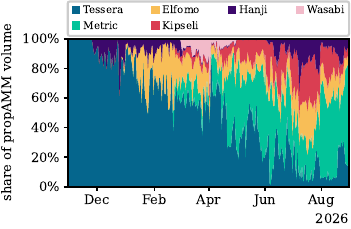}\vspace{-8pt}
  \caption{Base: share by propAMM}
\end{subfigure}\\[4pt]
\begin{subfigure}[b]{2.0in}\centering
  \includegraphics[scale=1]{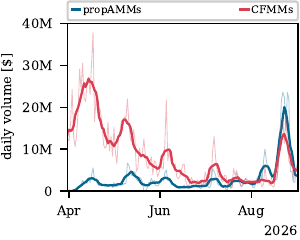}\vspace{-8pt}
  \caption{Monad MON/USDC: volume}
\end{subfigure}\hspace{10pt}%
\begin{subfigure}[b]{2.34in}\centering
  \includegraphics[scale=1]{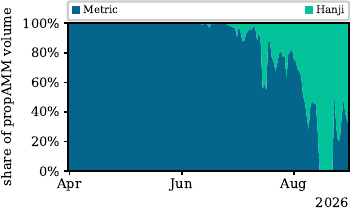}\vspace{-8pt}
  \caption{Monad: share by propAMM}
\end{subfigure}\vspace{-7pt}
\caption{Daily volume of propAMMs and AMM pools (left) and share of propAMM volume by venue (right), for SOL/USDC on Solana, ETH/USDC on Base and MON/USDC on Monad.}
\label{fig:volume_daily}\vspace{-20pt}
\end{figure}
\section{The PropAMM Landscape}

First, we look at the propAMM landscape on Solana, Base and Monad in Figure~\ref{fig:volume_daily}. For the main traded pair per chain, we show the daily propAMM volume in comparison to the AMM volumes, as well as the volume share of propAMMs. 

On Solana, average daily propAMM volume is \$1.12~billion, while it is \$292~million on AMMs for the SOL/USDC pair. Both decrease significantly throughout our measurement period. While propAMM volume is \$1.82~billion in the first 90 days, it falls to \$530~million in the last 90 days. We see a similar, though less extreme, pattern for AMM volume that falls from \$564~million to \$180~million. Thus, the volume for the SOL/USDC pair is far higher on propAMMs than on AMMs throughout, though the ratio falls from $3.2$ to $3.0$. 

In regard to the share of volume across propAMMs, we observe a significant concentration. In the first five months of our data, HumidiFi dominates the market with more than 50\% of the daily volume on average. The decrease in volume share of HumidiFi coincides with BisonFi entering the market. From the start of 2026, BisonFi dominated with a $40.3\%$ share of daily volume on average. Tessera also has a notable share of daily volume throughout our measurement period with $16.5\%$ on average, followed by SolFi V2 with $7.4\%$, while every remaining propAMM stays below $7\%$. 

Turning to Base, we start by noting that the propAMM volume is significantly lower than on Solana. Importantly, propAMMs only launch on Base two months into our measurements, on 30 October 2025, and their volume shows a steady increase from then, starting with \$9.5\,M per day on average in the first three months of their existence and reaching \$29.6\,M per day on average in the last three months of our measurement period. Over the same period, the proportion of ETH/USDC volume on propAMMs in comparison to AMMs increases from $2.4\%$ to $10.9\%$. In terms of the share of volume across propAMMs, Tessera, the first propAMM to launch on Base and also one of the major propAMMs on Solana, initially dominates. The share of volume by Tessera continuously falls as competitors start to launch. Tessera, however, remains the largest propAMM on Base in terms of total volume across our measurement period. 

Last, we turn to Monad. Monad launched nearly three months into our measurement period, on 24 November 2025, and the first propAMM launched in early April. Thus, we only have propAMM and AMM volume starting from then.\footnote{Note that there are two other propAMMs with significant volume, Clober Vault and Poe, on Monad. They are outside our scope, as they are not proprietary and their liquidity provision is permissionless~\cite{lfj_poe,lfj_poe_intro,clober_earn}.} We not only see propAMM volume on Monad picking up throughout our measurement period (\$2.1\,M in the first three months vs.\ \$3.9\,M in the last three months of our measurement period) but also see the share of MON/USDC volume on propAMMs vs.\ AMMs increase from $12.5\%$ to $45.7\%$ over the same period.\footnote{Note that on Monad, a significant share of MON/USDC volume is on central limit order books such as Kuru which are outside of our study~\cite{kuru_io,defillama_monad_dexs}.} 

\section{The PropAMM Edge: Markouts against AMMs}\label{sec:markouts}
Liquidity providers in AMMs are passive, i.e., their quotes only move once a trade has occurred, which exposes them to adverse selection~\cite{milionis2026automatedmarketmakinglossversusrebalancing,milionis2023automated}. PropAMMs emerged in part as a response. A single market maker supplies the inventory from its own vault and updates quotes on chain without a trade, cheaply enough to reprice at a high frequency. Thus, the market maker can move its price before a stale quote is picked off, provided its update is ordered ahead of the incoming trade. We test whether this advantage materializes. We measure markouts, i.e., the profit or loss on a fill evaluated against the reference price at a fixed horizon after execution, for both liquidity providers on AMMs and propAMM operators.

We evaluate every fill against the Bybit microprice~\cite{databento_microprice}, i.e., the weighted mid-price. We compute the microprice on the USDT pair, and convert it to USDC with the Bybit USDCUSDT mid-price.\footnote{Note that we compute the microprice on the USDT pair, as it has a significantly larger volume on Bybit for ETH, SOL and MON.}
For each fill, we compute the markout at horizon $\tau$ as the relative difference between the microprice $\tau$ seconds after execution and the execution price. We report the notional-weighted average across all fills of a propAMM. Positive values indicate that the propAMM gained. 

\begin{figure}[t]\centering\vspace{-10pt}
\begin{subfigure}[b]{0.32\linewidth}\centering\includegraphics[scale=1]{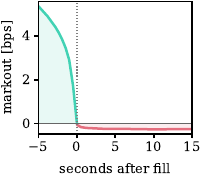}\caption{Solana}\end{subfigure}\hfill
\begin{subfigure}[b]{0.32\linewidth}\centering\includegraphics[scale=1]{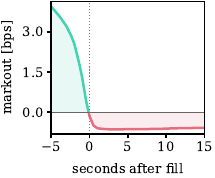}\caption{Base}\end{subfigure}\hfill
\begin{subfigure}[b]{0.32\linewidth}\centering\includegraphics[scale=1]{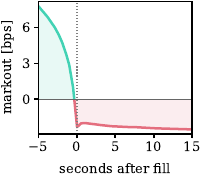}\caption{Monad}\end{subfigure}\vspace{-6pt}
\caption{Notional-weighted mean gross maker markout of AMM pools against the Bybit microprice $t$ seconds after the fill, for SOL/USDC on Solana, WETH/USDC on Base and WMON/USDC on Monad.}
\label{fig:markouts_dex}\vspace{-8pt}
\end{figure}

\subsection{AMM Markouts}

Figure~\ref{fig:markouts_dex} plots the markouts of AMMs on Solana, Base and Monad. While the magnitude of the markouts for AMMs differs across the three chains, the shape is identical. On all three chains the markouts are strongly positive ahead of the fill, at $5.34$\,bps on Solana, $3.95$\,bps on Base and $7.76$\,bps on Monad 5\,s before; they steeply turn negative around the fill. AMMs only reprice once a trade has occurred. The loss 2\,s after the fill is $0.22$\,bps, $0.62$\,bps, and $2.06$\,bps on Solana, Base and Monad, respectively.

The shape of the markout curves is consistent with liquidity providers being picked off. A move in the reference price leaves them stale, and the arbitrageur who corrects the mispricing executes at the old price, subjecting the liquidity providers to LVR. The difference in magnitude in part reflects the volatility of the respective tokens, i.e., $60\%$ for ETH, $66\%$ for SOL and $123\%$ for MON over our measurement period, since LVR scales with the variance of the price and is more sensitive to proportional changes in volatility than to block times~\cite{milionis2026automatedmarketmakinglossversusrebalancing}.
\subsection{PropAMM Markouts}
\begin{figure}[b]\centering\vspace{-10pt}
\begin{subfigure}[b]{0.32\linewidth}\centering
  \includegraphics[scale=1]{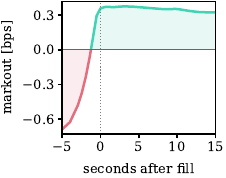}\vspace{-4pt}
  \caption{Solana}
\end{subfigure}\hfill%
\begin{subfigure}[b]{0.32\linewidth}\centering
  \includegraphics[scale=1]{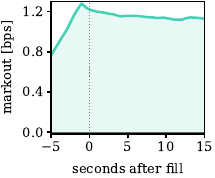}\vspace{-4pt}
  \caption{Base}
\end{subfigure}\hfill%
\begin{subfigure}[b]{0.32\linewidth}\centering
  \includegraphics[scale=1]{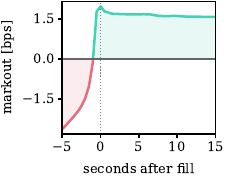}\vspace{-4pt}
  \caption{Monad}
\end{subfigure}\vspace{-6pt}
\caption{Notional-weighted mean gross maker markout of propAMMs
against the Bybit microprice $t$ seconds after the fill, for SOL/USDC on Solana, WETH/USDC on Base and WMON/USDC on Monad.}
\label{fig:markouts_prop}\vspace{-12pt}
\end{figure}

Next, we move to propAMMs. Figure~\ref{fig:markouts_prop} pools every propAMM on each chain, and the curves invert the AMM pattern of Figure~\ref{fig:markouts_dex}. An AMM is in the money before the fill and loses after it, whereas a propAMM is out of the money before the fill and earns after it. On Solana and Monad the markout starts at $-0.69$ and $-2.65$\,bps $5$\,s before the fill, rises steeply into $t=0$, and stays positive for the rest of the horizon. Fills do not arrive at random points in time but follow moves in the reference price, and the pre-fill leg traces those moves. The difference between the two venue types is what happens to the quote during the move. An AMM waits for a trade, so the arbitrageur who corrects the mispricing executes at the stale price. A propAMM reprices as the reference moves and fills at the new quote, so it is not the one being picked off.

Two seconds after the fill propAMMs earn $0.37$\,bps on Solana, $1.19$\,bps on Base and $1.69$\,bps on Monad, while AMMs lose $0.22$, $0.62$ and $2.06$\,bps. Base is the exception to the shape, as its markout never turns negative and propAMMs there are ahead of the reference price even $5$\,s before the fill. Appendix~\ref{app:markout_prop_base_mon} provides a per-venue breakdown for the biggest propAMMs on Solana, Base and Monad.

\section{Where the Edge Comes From}\label{sec:edge}
Next, we examine how propAMM design supports competitive quotes while managing exposure to adverse selection. We base our analysis on both measurements and insights from Tessera's implementation on Base.

\subsection{Frequent and Inexpensive Quote Updates }

As highlighted previously~\cite{lostin2025propamm}, even though propAMMs' quotes can be updated via trades indirectly, they also update their quotes by sending cheap update transactions, e.g., by changing the anchor price $p$. Such an update is far cheaper than a swap. On Solana, a median update consumes between $485$ and $676$ CU against at least $16{,}938$ for a swap, i.e., more than an order of magnitude less. Similarly, on Base a median update costs between $26{,}774$ and $32{,}606$ gas while simulated gas consumption for single swap transactions against the 31st of August snapshot ranges from $201{,}243$ to $402{,}881$. On Monad, the median gas consumption for an update ranges from $33{,}428$ to $50{,}839$ and median gas limits from $45{,}000$ to $70{,}000$. For single swap transactions, simulated gas consumption lies around the lower bound $162{,}503$ with access list and $330{,}103$ without access list for Metric.\footnote{Note that Monad charges on \texttt{gasLimit}~\cite{categorylabs2025monadspec}. The unused fraction of gas is around $28.32\%$ for updates against $38\%$ in a sample of non-update transactions, i.e., propAMM updates set \texttt{gasLimit} more tightly.}
Note that on both EVM chains an update is $5$ to $15$ times cheaper than a swap, and that the $21{,}000$ gas floor every transaction pays accounts for much of what an update costs.

Across the three chains, transactions are by default ordered by priority fee per unit of compute. PropAMMs exploit the difference in transaction size by bidding cheaply for an early block position. On Solana, they pay between $304$K and $2.2$M micro-lamports per CU on an update, against $5.6$K to $22$K on a trade. Interestingly, this does not hold across all updates, as $32\%$ of Tessera's updates and $27\%$ of BisonFi's pay neither a priority fee nor a tip. We observe a similar pattern on Base. Tessera pays $2.37$ times its base fee to land a price update, but only $0.18$ times for a book update and $0.77$ times for a default-fee update, i.e., it pays for position only when the update is time critical. Additionally, as opposed to its Solana deployment, only $0{.}113\%$ of its update transactions do not pay any priority fee. On Monad, all recorded updates pay a non-zero priority fee for Hanji and Metric with median effective priority fees ranging from $2$ to $10.5$ MON-gwei per gas, whereas for trades these numbers range from $3$ to $40$.

\parhead{Tessera Implementation Insight: Quote Freshness.}\label{par:quote_freshness} Tessera's Base implementation further guards against the case where the operator fails to update in time. Each price update records a reference block alongside the anchor price. A trader executing against it receives no freshness penalty for a short grace period, an increasingly worse price thereafter, and no fill at all once the quote expires. 

The low cost of updates allows propAMMs to update quotes frequently. For Tessera on Base, we measure the correlation between successful price-update counts and ETH realized volatility in non-overlapping five-minute windows. Volatility is estimated from five-second log returns. We compute the Pearson correlation between raw update counts and raw volatility after removing calendar-day and hour-of-day effects from both series, obtaining $0.56$. Note that we exclude default-fee and other non-price updates, as well as unsuccessful updates. Thus, Tessera updates its prices more often during times of high volatility, when prices move faster.

\subsection{Pricing by Counterparty}\label{sec:counterparty}

\begin{wraptable}{r}{0.52\textwidth}\centering\setlength{\tabcolsep}{4pt}\footnotesize
\vspace{-\intextsep}\vspace{-20pt}
\resizebox{\linewidth}{!}{\begin{tabular}{lrrrr}\toprule
 & \multicolumn{2}{c}{median spread} & \multicolumn{2}{c}{paired} \\
\cmidrule(lr){2-3}\cmidrule(ll){4-5}
propAMM & aggr. & other & wider & premium \\ \midrule
HumidiFi & 0.09 & 0.86 & 77\% & +0.40$^{***}$ \\
BisonFi & 0.71 & 1.01 & 76\% & +0.10$^{***}$ \\
Tessera & 0.90 & 0.93 & 68\% & +0.10$^{***}$ \\
SolFi V2 & 3.88 & 4.90 & 82\% & +1.58$^{***}$ \\
GoonFi & 0.78 & 4.39 & 83\% & +3.59$^{***}$ \\
ZeroFi & 1.52 & 1.50 & 41\% & -0.05 \\
\bottomrule\end{tabular}}\vspace{-4pt}
\caption{Realized round-trip spread (bps) of the six largest Solana propAMMs on SOL/USDC, by whether an aggregator, e.g., Jupiter, routed the fill. Round trips are consecutive opposite-side fills of similar size on one pool with no intervening quote update, with both legs of the same class, compared within (pool, day, size bucket): \emph{wider} is the share of cells where the non-aggregator median is larger, \emph{premium} the median within-cell difference. Sign test: $^{*}p<0.05$, $^{**}p<0.01$, $^{***}p<0.001$.}
\label{tab:counterparty}\vspace{-14pt}\end{wraptable}

To protect against adverse selection, propAMMs are said to classify their order flow and quote different prices depending on how the trade is routed to them~\cite{latif2026propamms}.

\parhead{Solana.} We measure the realized spread for the top six Solana propAMMs for aggregator swaps vs.\ non-aggregator swaps. Whenever the same pool fills two trades on \emph{opposite} sides within 25 slots, of a similar size (within a factor of two), and without an update from the propAMM in between, the two fills form a round trip. Its realized spread is the relative gap between the two prices. Notice that for all propAMMs but ZeroFi, non-aggregator flow receives the wider spread, and the direction is highly statistically significant: it holds in $68$--$83\%$ of matched cells, with an exact sign test yielding $p<0.001$. The size of the gap, however, varies widely. GoonFi charges non-aggregator flow $3.59$\,bps more and SolFi V2 $1.58$\,bps more, while for BisonFi and Tessera the difference is only $0.10$\,bps. 

\parhead{Tessera Implementation Insight: Fee Components.} The Tessera implementation on Base charges six fee components: base, age, priority, mode, default and address. The last two depend on who is trading, i.e., on both the transaction's sender and the contract calling Tessera's interface. Addresses fall into four categories: (1) blacklisted addresses cannot trade, (2) whitelisted (i.e., default exempt) addresses do not pay the default fee, (3) penalized addresses pay an additional fee, and (4) remaining addresses. An address may also be penalized if its prefix or bytecode matches a known pattern. Appendix~\ref{app:tessera_class} gives the details.

\begin{wrapfigure}{r}{0.56\linewidth}\centering\vspace{-10pt}
\includegraphics[scale=1]{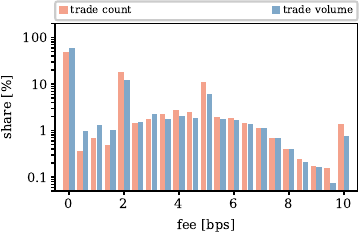}\vspace{-6pt}
\caption{Tessera's total fee across its WETH/USDC trades on Base, by trade count and by volume, with fees of $10$\,bps and above in the last bar.}
\label{fig:tessera_fee_hist}\vspace{-10pt}
\end{wrapfigure}

For each swap we measure the marginal cost to the trader, i.e., the half spread taking the mid price derived from the top of the book marginal executable prices, \textit{nominal} before fees and \textit{effective} after (Appendix \ref{app:tessera_fee}). Volume-weighted, the nominal marginal cost is $0.44$\,bps and the effective marginal cost $1.95$\,bps, so the fee, not the book, accounts for most of the marginal cost. The fee is not spread evenly across trades but concentrated on a few discrete levels (Figure~\ref{fig:tessera_fee_hist}). Tessera charges no penalties on $42.1\%$ of trades, carrying $47.4\%$ of volume, and a further $7.0\%$ of trades, or $13.0\%$ of volume, pay only the smallest unit the contract charges, $0.01$\,bps. Two spikes account for most of the rest, at $[1.75,2.25)$\,bps for $18.0\%$ of trades and $[4.75,5.25)$\,bps for $11.1\%$. Trades accounting for $60.4\%$ of total volume pay at most $0.01$\,bps, i.e., low penalty trades are larger on average.

Tessera supports six penalty components, five of which are nonzero in our sample. Priority is the fee component charged most often, on $42.7\%$ of all trades and as the sole charge on $34.4\%$, and it explains both spikes in the histogram. Priority is the sole component for 99.91\% and 99.41\% of swaps paying exactly 2 and 5\,bps, respectively. Default is charged on $22.7\%$ of trades, and is the sole charge on $14.6\%$, yet it collects most of the fee revenue (Appendix \ref{app:tessera_fee}), at $56.2\%$ against priority's $34.9\%$, so it lands on the largest trades. Address, age and mode are each the sole charge on under $0.4\%$ of trades and together account for under $9\%$ of revenue. Address classification differentiates traders. Swaps from currently blacklisted addresses executed at an effective marginal cost of $2.62$\,bps before blacklisting, address-penalized swaps at $4.96$\,bps, whitelisted swaps at $0.78$\,bps and all remaining swaps at $2.46$\,bps.

\subsection{Arbitrage Flow}
PropAMMs allow operators to update on-chain quotes directly in response to external market prices. Price discrepancies between a propAMM and an AMM or another propAMM can create atomic arbitrage opportunities, with each venue supplying a leg of the trade. We separate fills by the transaction they occur in, i.e., single fills, propAMM--propAMM arbitrage, propAMM--DEX arbitrage and other multi-fill transactions. Appendix~\ref{app:mev} details arbitrage classification.

\begin{wraptable}{r}{0.6\linewidth}\vspace{-14pt}
\centering\setlength{\tabcolsep}{4pt}\footnotesize
\resizebox{\linewidth}{!}{%
\begin{tabular}{@{}lrrrrrrrr@{}}\toprule
 & \multicolumn{2}{c}{single} & \multicolumn{2}{c}{prop--prop} & \multicolumn{2}{c}{prop--DEX} & \multicolumn{2}{c}{other} \\
\cmidrule(lr){2-3}\cmidrule(lr){4-5}\cmidrule(lr){6-7}\cmidrule(l){8-9}
 & share & 2\,s & share & 2\,s & share & 2\,s & share & 2\,s \\ \midrule
Solana & 50.5\% & +0.41 & 16.6\% & $-$0.21 & 4.8\% & +0.03 & 28.1\% & +0.96 \\
Base   & 29.7\% & +0.92 & 4.4\% & $-$0.22 & 13.3\% & +0.02 & 52.6\% & +1.82 \\
Monad  & 63.7\% & +1.44 & 4.7\% & +0.68 & 6.4\% & +1.55 & 25.2\% & +2.57 \\
\bottomrule\end{tabular}}\vspace{-8pt}
\caption{PropAMM fills by route class: share of notional and notional-weighted markout (bps) at 2\,s. Single means one decoded propAMM fill and no other swap. Prop--prop and prop--DEX denote atomic arbitrage against a second proprietary pool and against a public pool, respectively. Other covers all remaining transactions, such as split orders and unresolved legs. Shares are per transaction, counting the propAMM's own fill to avoid double counting. Markouts weight every fill by its notional, as in Section~\ref{sec:markouts}.\protect\footnotemark{}}
\label{tab:route_compact}\vspace{-4pt}
\end{wraptable}
\footnotetext{Appendix~\ref{app:mev} lists the companion venues we decode on each chain, and those we recognise but cannot resolve.}

Table~\ref{tab:route_compact} splits propAMM fills by the transaction they occur in. Transactions with a single propAMM swap are the largest class on Solana and Monad, and have positive markouts on every chain at 2\,s: $0.41$\,bps on Solana, $0.92$\,bps on Base and $1.44$\,bps on Monad. The remaining flow, i.e., order splitting across venues, longer routes and unresolved companion legs, carries $28.1\%$ of notional on Solana, $52.6\%$ on Base and $25.2\%$ on Monad.

PropAMM--DEX arbitrage is $5\%$ of notional on Solana, $13\%$ on Base and $6\%$ on Monad. These legs are profitable on Monad, earning $1.55$\,bps, and close to flat on Solana and Base, at $0.03$ and $0.02$\,bps. Before a propAMM--DEX arbitrage, the reference price moves by about $4.9$\,bps on Solana, about $1.9$\,bps on Base and about $7.3$\,bps on Monad (Figure~\ref{fig:markouts_prop_dex_venues} in Appendix~\ref{app:route_class}). The propAMM reprices in response, whereas the AMM does not. An arbitrageur then trades the propAMM's updated quote against the stale pool. The propAMM thus captures its spread against the stale pool.

PropAMM--propAMM arbitrage accounts for $16.6\%$ of notional on Solana, but under $5\%$ on Base and Monad. In these transactions, the arbitrageur trades between two proprietary pools whose quotes briefly disagree.

\begin{figure}[t]\vspace{-10pt}\centering
\begin{subfigure}[t]{0.32\linewidth}\centering
\includegraphics[width=\linewidth]{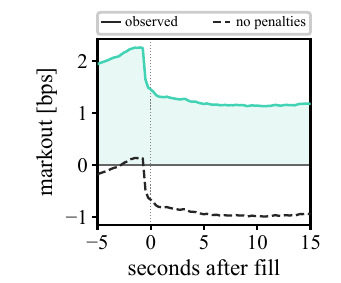}
\caption{single fills}\label{fig:tessera_mev_single}
\end{subfigure}\hfill
\begin{subfigure}[t]{0.32\linewidth}\centering
\includegraphics[width=\linewidth]{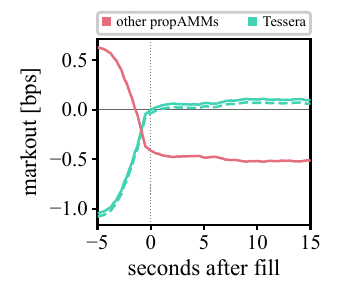}
\caption{propAMM--propAMM}\label{fig:tessera_mev_propamm}
\end{subfigure}\hfill
\begin{subfigure}[t]{0.32\linewidth}\centering
\includegraphics[width=\linewidth]{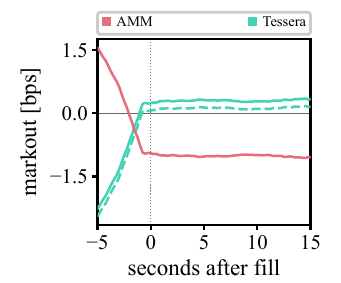}
\caption{propAMM--AMM}\label{fig:tessera_mev_AMM}
\end{subfigure}\vspace{-4pt}
\caption{Tessera ETH/USDC markouts on Base by transaction type, observed and at the unpenalized quote. Shares of total Tessera volume: (a) 20.70\% (\$895.03M), (b) 8.04\% (\$347.72M); (c) 13.62\% (\$588.84M).}
\label{fig:tessera_mevn}
\end{figure}

\parhead{Penalties and Atomic Arbitrage on Tessera.} Recall that propAMMs apply penalties as described in Section~\ref{par:quote_freshness}. For Tessera's ETH/USDC pool on Base, where we decode the penalty of every fill, we can also recompute the markout the trader would have received at the unpenalized quote. In the following, we look at the various flows received by Tessera with the aim of understanding how the penalties affect their markouts. Figure~\ref{fig:tessera_mevn} shows the markouts, with and without penalties, for propAMM--AMM arbitrage, propAMM--propAMM arbitrage and single fills from the perspective of Tessera. 

On single fills, Tessera earns $1.29$\,bps at 2\,s, but would have lost about $1$\,bps had it filled without penalties (Figure~\ref{fig:tessera_mev_single}). These fills also pay a volume-weighted effective cost of $2.58$\,bps, $32.4\%$ more than the pool as a whole. Some transactions in this category are likely attempted CEX--DEX arbitrage: the swaps in this category target the side of the Tessera book which the reference price move exposes to adverse selection in the second before the fill. Instead of taking a loss on a stale quote, however, Tessera appears to identify that it is being adversely selected and charges penalties high enough to turn a profit on single fills.

For propAMM--AMM arbitrage legs (Figure~\ref{fig:tessera_mev_AMM}), Tessera's observed markout sits slightly above zero. The reference price moves by more than $1$\,bps toward Tessera's execution price in the seconds before the fill, i.e., these fills follow a reference move. Tessera therefore executes close to fair value and, while it does not have negative markouts, it also does not earn a lot on these legs.

Figure~\ref{fig:tessera_mev_propamm} looks at propAMM--propAMM arbitrage through Tessera, which makes up $8.04\%$ of the total Tessera ETH/USDC volume and \$347.72~million of volume on the other propAMMs' ETH/USDC pools. Here the two legs look nothing alike. Tessera's markout stays slightly above zero with or without penalties. The other propAMM is in the money before the fill and down about $0.5$\,bps after it, i.e., the same shape we see for AMMs in Figure~\ref{fig:markouts_dex}. Tessera is the leg that has repriced, and the other propAMM the one that has not.

Overall, the mechanics of propAMM--AMM/propAMM atomic arbitrage cases resemble those of traditional CEX--DEX arbitrage: a propAMM is keeping up with the reference price, while a different AMM/propAMM pool is stale with respect to the reference price. This creates opportunities for arbitrageurs. They can arbitrage the stale AMM/propAMM against the CEX or arbitrage it against the non-stale propAMM, executing at lower margins but executing atomically, mitigating the risk of failure of the second (i.e., CEX) leg.

\subsection{Quoting Accuracy}\label{sec:quoting_accuracy} 
The mechanisms examined so far protect a propAMM against adverse selection, which is ordinary market making. We now turn to a less benign source of the edge, i.e., the gap between the price a propAMM advertises and the price at which it actually fills.

\begin{figure}[t]\centering\vspace{-10pt}
\begin{subfigure}[t]{0.49\linewidth}\centering
\includegraphics[width=\linewidth]{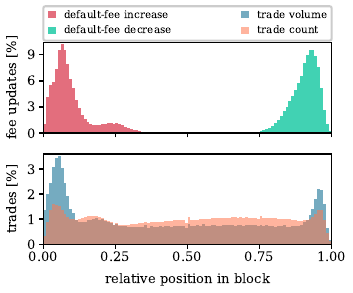}\vspace{-4pt}
\caption{Default-fee updates and trades}\label{fig:tessera_fee_position}
\end{subfigure}\hfill
\begin{subfigure}[t]{0.49\linewidth}\centering
\includegraphics[width=\linewidth]{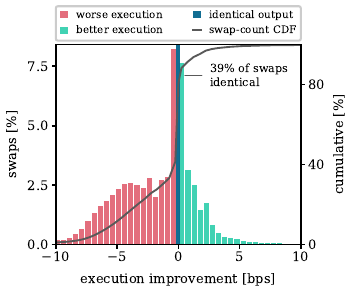}\vspace{-4pt}
\caption{Execution quality}\label{fig:tessera_exec_quality}
\end{subfigure}\vspace{-4pt}
\caption{Tessera execution quality and default-fee updates. (a) Default-fee updates and trades by relative transaction position in the block ($0$ = first, $1$ = last). (b) Swaps by the difference between executed and default output.}\vspace{-10pt}
\label{fig:tessera_bins}
\end{figure}

\parhead{Tessera Implementation Insight: Spoofing.} PropAMMs are said to quote better prices at the end of the block, i.e., in the state that aggregators use to compute their routes off-chain. They thereby \textit{spoof} aggregators, which route to the venue based on the price they observe but receive a worse one at the start of the next block. Figure~\ref{fig:tessera_fee_position} shows this pattern for Tessera on Base. Default-fee decreases concentrate in the final tenth of the block and increases in the first tenth. Thus, the effective spread is wider right after the fee increase early in the block, and narrower at the very end, i.e., in exactly the state aggregators read.

Interestingly, volume peaks both at the very start and at the very end of the block, and more sharply than trade count, i.e., trades at the block edges are also larger than average. Notably, the volume peak at the start arrives just before the fee increase, and the peak at the end just after the fee decrease. Large traders thus appear to be aware of Tessera's pattern and to time their trades into the low-fee window spanning the end of one block and the start of the next.

Figure~\ref{fig:tessera_exec_quality} summarizes the effect. Only 39\% of trades receive the same execution as they would have based on the quote at the end of the previous block. On average, however, swaps receive $1.08$\,bps worse execution, and $0.56$\,bps worse when weighted by volume, likely because volume concentrates in the low-fee part of the block. Recall that Tessera charges an additional penalty on trades paying a high priority fee (Section~\ref{sec:counterparty}), likely to protect against trades that try to execute ahead of its repricing at the start of the block.

Spoofing results in traders receiving, on average, a worse price than the quote they saw, and sometimes a better one, due to price changes between blocks as well as opposite-direction swaps which might reintroduce liquidity to the books as visualized in Figure~\ref{fig:tess_exec} (Appendix~\ref{app:tessera_q_e}). We note that a similar gap between quoted and executed prices has been reported on Solana~\cite{0x2026propamm}, though we lack the historical quote data to corroborate it.

\section{Retail Execution on PropAMMs}\label{sec:retail}
The promise of propAMMs is not only that they are able to avoid LVR through active quoting, but also that as a result they can offer users good prices. Thus, we next isolate retail flow. For each fill we measure the absolute move of the Bybit weighted mid between $5$\,s before and $1$\,s after execution, and call a fill \emph{quiet} when that move stays below $1$\,bps and \emph{moving} otherwise. Quiet fills arrive independently of what the market is doing and carry no information the maker has to price against, which is what we expect of retail.

\begin{figure}[t]\centering\vspace{-6pt}
\begin{subfigure}[b]{0.32\linewidth}\centering
  \includegraphics[scale=1]{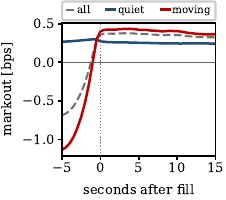}\vspace{-4pt}
  \caption{Solana}
\end{subfigure}\hfill%
\begin{subfigure}[b]{0.32\linewidth}\centering
  \includegraphics[scale=1]{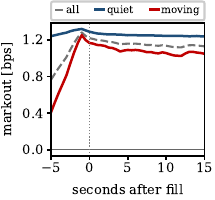}\vspace{-4pt}
  \caption{Base}
\end{subfigure}\hfill%
\begin{subfigure}[b]{0.32\linewidth}\centering
  \includegraphics[scale=1]{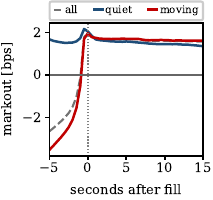}\vspace{-4pt}
  \caption{Monad}
\end{subfigure}\vspace{-4pt}
\caption{PropAMM markouts split by reference price movement from $5$\,s before to $1$\,s after the fill: quiet ($<1$\,bps) against moving ($\geq 1$\,bps). Quiet fills carry $32\%$ of notional on Solana, $43\%$ on Base and $17\%$ on Monad (Hanji and Metric).}
\label{fig:quiet_prop}
\end{figure}

Quiet flow pays $0.26$\,bps on a Solana propAMM, $1.26$\,bps on Base and $1.62$\,bps on Monad (Figure~\ref{fig:quiet_prop}). Interestingly, quiet and moving fills differ by at most $0.16$\,bps at $2$\,s on every chain. The two classes differ drastically before the fill, where the steep drift of the pooled curves comes entirely from moving flow and the propAMM repricing ahead of it. On quiet flow the markout is flat across the whole horizon. Thus, its level approximates what a retail trader pays relative to the reference price.

Compared with AMMs, propAMMs take a larger share of retail flow (Appendix~\ref{app:retail_AMM}). Quiet fills carry $32\%$ of propAMM notional on Solana, $43\%$ on Base and $17\%$ on Monad, against $9\%$, $16\%$ and $8\%$ on AMMs. The half-spread retail pays, i.e., the distance from the reference price to the execution price, is $0.26$\,bps on a Solana propAMM against $2.59$\,bps on a Solana AMM, $1.62$ against $8.60$\,bps on Monad and $1.26$ against $1.38$\,bps on Base. Thus, propAMMs give users a better price than AMMs do, and the edge their operators hold over AMM liquidity providers does not come from retail.

\section{Concluding Discussion}
PropAMMs account for a large share of DEX volume across all three chains in our study. We show that their operators have an edge over passive liquidity providers on AMMs and detail where it comes from. PropAMMs update prices cheaply and frequently without a trade, charge by counterparty, do not lose to arbitrage against other on-chain venues, and spoof, i.e., quote better prices than a trade receives. The edge does not come out of retail, which pays less on a propAMM than on an AMM and makes up a larger share of propAMM volume.

Currently, users bear the cost of spoofing. Aggregators route on stale quotes read from the previous block. As a result, only $39\%$ of swaps on Tessera execute at the quoted price, and the average swap receives $1.08$\,bps less. An on-chain router that queries each venue inside the swap transaction would remove this gap and route each trade on the price it actually receives.

\bibliographystyle{splncs04}
\bibliography{refs}

\appendix
\section{Data Collection}\label{app:data_collection}
Our measurement spans $1$ September $2025$ to $31$ August $2026$, with Monad data beginning at its mainnet launch on $24$ November $2025$.

\subsection{Swaps}

On Solana we stream the whole-epoch CAR archives of Old Faithful~\cite{old_faithful} for epochs $826$ to $1026$, and record each transaction's position in its block, its fee payer and its full fee stack. We track $14$ propAMM programs across $13$ venues with SOL/USDC volume, i.e., HumidiFi, BisonFi, Tessera, SolFi, SolFi V2, GoonFi, GoonFi V2, ZeroFi, Aquifer, Scorch, Quantum, AlphaQ and Obric V2. These give $448{,}582{,}796$ trades, and we record each venue's quote updates alongside its trades. Update counts are an estimate: we classify a transaction as a trade if it carries a propAMM fill and otherwise as an update if it invokes the propAMM program without swapping, so any non-swap interaction is counted as an update and we cannot observe whether a given update moved the venue's quote. Our AMM coverage spans Orca Whirlpool, Raydium CLMM, Raydium AMM v4, Raydium CPMM and Meteora DLMM, i.e., $144{,}106{,}459$ SOL/USDC fills. Note that update counts are an estimate, as we count every non-swap propAMM transaction as an update.

On Base and Monad we collect swaps by decoding the respective log events, along with the transaction index, log index and sender. On Base our coverage spans Aerodrome Slipstream, Uniswap v3, PancakeSwap v3, Uniswap v4, Aerodrome v1 and Uniswap v2, together with a tail of v2- and v3-style forks, giving $148{,}962{,}685$ ETH/USDC swaps across $240$ pools.  On Monad we cover Uniswap v4, in both its native-MON and its WMON pools, PancakeSwap v3, Uniswap v3, LFJ Liquidity Book v2.2 and Uniswap v2, again with a tail of v3-style forks, giving $15{,}659{,}039$
MON/USDC swaps across $115$ pools.

\subsection{Markout Timestamps}
\parhead{Solana.} Solana block timestamps have one-second resolution, which cannot separate the two to three slots that fall in the same second at a $400$\,ms slot time. The vote transactions carry a finer signal. Validators vote on a slot in one of the subsequent blocks, and each vote is stamped with the voting validator's own clock time. In most slots, a single second carries around $99\%$ of the votes. When a slot falls near a second boundary, the votes split between the two adjacent seconds, and the size of that split locates the boundary within the slot. Calibrating against Anza's validator metrics~\cite{anza_metrics}, we find that the votes for a slot are cast within a window about $0.22$\,s wide. This width translates the observed split into a sub-second offset, which we use to place the slot. These boundary slots serve as anchors. The slots between two anchors are placed by linear interpolation in slot number, and those past the last anchor by the local slot cadence.

\begin{wrapfigure}{r}{0.45\textwidth}
    \centering
    \vspace{-2pt}
    \includegraphics[width=\linewidth]{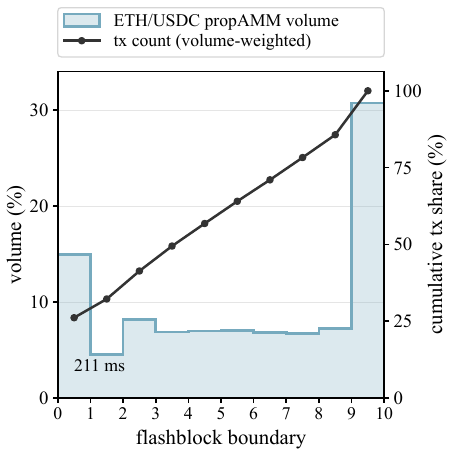}
    \vspace{-6pt}
    \caption{ETH/USDC propAMM volume by Flashblock.
    The line shows cumulative transaction shares,
    averaged across blocks using propAMM volume weights.}
    \label{fig:base-markout-timestamps}
    \vspace{-10pt}
\end{wrapfigure}

\parhead{Base.} We use recorded Flashblock preconfirmations to design a two-point timestamp approximation for trades without observed Flashblock timestamps. In the overlapping trade and Flashblock sample, ETH/USDC propAMM volume is concentrated in FB1 and FB10, which together account for 45.71\% of volume (Figure~\ref{fig:base-markout-timestamps}).\footnote{FB0 contains only a system transaction and is omitted from the figure.}

For each block, we compute the cumulative transaction fraction through FB1. Averaging these fractions with propAMM ETH/USDC volume weights gives roughly 26\%. Additionally, the median FB1 receipt time is around 200\,ms after the preceding block's timestamp. We therefore assign trades in the first 26\% of a block's transaction positions the timestamp $b-1.8$\,s, and remaining trades the timestamp $b$, where $b$ denotes the current block's timestamp.  This coarse approximation captures early-block execution without separately modeling each Flashblock, whose receipt times and cumulative transaction fractions have high variance across blocks. It retains timing uncertainty, particularly for trades in intermediate Flashblocks.

\parhead{Monad.} For Monad markouts, we use higher-resolution MonadBFT timestamps, i.e., consensus-layer timestamps.

\subsection{Reference Prices}
We evaluate every fill against a Bybit top-of-book reference. We download Bybit's public
$200$-level spot orderbook archive and record the best bid and ask with
their sizes, the mid, and the size-weighted mid
$(\text{bid}\cdot q_{\text{ask}} + \text{ask}\cdot q_{\text{bid}})/(q_{\text{bid}}+q_{\text{ask}})$. We collect SOLUSDT,
ETHUSDT, MONUSDT and USDCUSDT for 1 September 2025 to 31 August 2026. Note that we compute the reference on the USDT pair, which is substantially more liquid than the USDC
pair for all three base assets, and convert to USDC by dividing by the Bybit USDCUSDT mid.

\section{PropAMM MEV Detection Heuristics}
\label{app:mev}
\parhead{Atomic Arbitrage.} We collect all swap logs for transactions involving the ETH/USDC Tessera pool and then employ a simple transaction filter by aggregating the trader's balances in a dictionary in the form $\{\texttt{token}_i\to \Delta_i\}$. We then mark the transaction as atomic arbitrage if $\exists i. \Delta_i >0\wedge \forall i. \Delta_i \geq 0$. Let $\mathcal A_\text{to}$ and $\mathcal A_\text{from}$ be the sets of \texttt{tx.to} and \texttt{tx.from} values involved in at least one transaction marked as atomic arbitrage in the previous step. For each set, we compute the share of its total Tessera ETH/USDC volume attributable to transactions marked as atomic arbitrage. This share is $92.9\%$ for $\mathcal{A}_{\mathrm{to}}$ and $99.1\%$ for $\mathcal{A}_{\mathrm{from}}$, with each denominator including all recorded Tessera ETH/USDC trades associated with that set. This gives us confidence that this simple classification is reasonable and accurate.

\parhead{Other Pools in the Transaction.} To assign a propAMM fill to a route class, we need to identify every other pool its transaction trades on. The classification is therefore only as complete as the set of venues we decode. On Solana, we decode Orca Whirlpool, Raydium CLMM, Raydium AMM v4, Raydium CPMM and Meteora DLMM. On Base, we decode Uniswap v2, v3 and v4, PancakeSwap v3, and Balancer v2 and v3, as well as Aerodrome v2, Algebra, the Angle Transmuter, DODO, Maverick v1 and v2, Pinto Well, Spark PSM3 and WOOFi. On Monad, we decode Uniswap v2-, v3- and v4-style pools, PancakeSwap v3, Balancer v3 and LFJ Liquidity Book. We also recognise swaps on the following venues, but cannot tell which pool they trade on or how much: on Solana, Pump.fun, Phoenix, OpenBook, Lifinity v2, Meteora Pools and Raydium Route, and on Base, Curve, Fluid DEX, PancakeSwap Infinity CL and Infinity Bin, Virtuals, 1inch and Kyber limit orders, and RFQ fills.

We count each swap we cannot decode as a separate pool. Since we cannot check which pair it trades or whether the trade balances, a transaction that contains one cannot be propAMM--propAMM or propAMM--DEX arbitrage, and we assign it to \emph{other}. One exception applies to 1inch and Kyber limit orders on Base. If a limit order only passes a propAMM swap through, i.e., the swap runs inside the order's fill and the fill moves nothing but that swap's amounts and fees, we do not count the order as a second venue. Finally, on Monad, most of Hanji's and Metric's propAMM--propAMM volume trades against permissionless venues outside our scope, e.g., Clober Vault and Poe.

\section{Tessera}\label{app:tessera}
In this section, we give further details regarding some aspects of Tessera.

\subsection{Marginal Cost and Fee Revenue}\label{app:tessera_fee}
For each swap, let $p_a$ and $p_b$ denote the first accessible ask and bid prices before execution penalties, expressed in USDC per ETH. Let $\boldsymbol{\phi}\in\mathbb{N}_0^6$ contain the applied penalty components in ppm.

\parhead{Marginal cost.} We define nominal and effective marginal costs as half-spreads normalized by their respective bid and ask midpoints, expressed in basis points:
\[
c_{\mathrm{nom}}
=10^4\frac{p_a-p_b}{p_a+p_b},
\qquad
c_{\mathrm{eff}}
=10^4\frac{p_a-p_b\cdot(1-10^{-6}\boldsymbol{\phi}^\top\mathbf 1)^2}
{p_a+p_b\cdot(1-10^{-6}\boldsymbol{\phi}^\top\mathbf 1)^2}
\]
\parhead{Fee Revenue.} For each swap, let $Y$ denote the exchanged USDC amount. Abstracting from integer rounding, we attribute USDC-equivalent fee revenue to the six components as
\[
\mathbf{r}
=
\frac{Y\boldsymbol{\phi}}
{10^6-\boldsymbol{\phi}^{\top}\mathbf{1}}
\]
For swaps receiving USDC, this reconstructs the USDC withheld by each component. For swaps receiving WETH, it values the withheld WETH at the swap's execution price.

\subsection{Swap Classification}\label{app:tessera_class}
Tessera combines manager-maintained address labels (ordinary, default-exempt, penalized, or blacklisted) with bytecode fingerprint checks. It checks the transaction origin and the public-interface caller. In either case, blacklisting rejects execution, exemption waives the default penalty, and penalization selects the address penalty.

If no explicit label selects the address penalty, the helper \href{https://basescan.org/address/0xfdb7fa3f95e47624b7423b48462564107aa4e684}{\texttt{0xfdb...684}} is used to fingerprint observable address and code features. These checks also apply to default-exempt addresses. A match selects the configured address penalty without changing stored labels.  
Next, we present two examples where the helper applies a penalty as a case study.

\parhead{Metric and Atomic Arbitrage.}
The helper matches on addresses starting with a \texttt{c0ffee} prefix. Transaction \href{https://basescan.org/tx/0x5dde4996c5d675f084976cf3cf8dece43f73f9c79ef4d3dd36db9128028fc62d}{\texttt{0x5dde...8fc62d}} originates from a \texttt{c0ffee}-prefixed address and completes an atomic three-swap cycle through Metric, Uniswap v3, and Tessera. Its route is in the form: USDC $\to$ WETH $\to$ VVV $\to$ USDC. Thus, the transaction performs on-chain arbitrage. Note that the execution invokes Metric's swap callback as well, another pattern the historical helper flags.

\parhead{Forwarding and CEX--DEX.}
Transaction \href{https://basescan.org/tx/0xac4f3dc15e4c011a71f02eeae55945f792433efce43ea353bfcbe02c66550c71}{\texttt{0xac4...0c71}} creates a 30-byte forwarding contract and uses it to exchange 128 WETH for approximately 303,365 USDC through Tessera. The transaction contains only one pool swap, which is consistent with the on-chain leg of CEX--DEX arbitrage. The markout of this swap is at $-0.510$ bps at 0s and falls further to $-5.288$ bps at 2s. Both the origin and the newly created contract have ordinary status, but the helper recognizes a forwarding pattern and selects a 175 ppm address penalty. Thus, Tessera classifies newly created accounts through a pattern in their bytecode without first assigning them an explicit address label. 

\parhead{Penalty Groups and Within-Block Position.}
We analyze trading flow by its applied penalties and position within the block. Let $t_i$ denote the estimated timestamp of swap $i$, $m(t)$ the reference price at time $t$, and $m_i(\tau)=m(t_i+\tau)$. We define the signed reference-price movement between horizons $\chi$ and $\tau$ as
\[I_i(\tau,\chi) = -10^4 d_i \frac{m_i(\tau)- m_i(\chi)}{m_i(\chi)} \]
where $d_i=1$ if Tessera buys ETH and $d_i=-1$ if it sells. Thus, by construction, positive values indicate reference-price movements in the trader's direction. Let $V_i$ denote the USDC-denominated volume of swap $i$. For each penalty component $k$, let $G_k$ contain swaps with a positive applied value of that component. These groups are non-disjoint, i.e., a swap with positive address and priority penalties will be included in both $G_{\text{address}}$ and $G_\text{priority}$. We also define an additional none group which consists of swaps with 0 total penalty.

To capture exposure to \textit{informative} flow, we measure the volume-weighted share of each group followed by a reference-price move exceeding $x$ bps in the direction benefiting the trader over the next 15 seconds, formally:

\[\gamma_{k}(x) = \frac{\sum_{i\in G_k}V_i\cdot \mathbf 1\{I_i(15,0) > x\}}{\sum_{i\in G_k}V_i}\]

\begin{figure}[t]
\centering
    \centering
    \includegraphics[width=\linewidth]{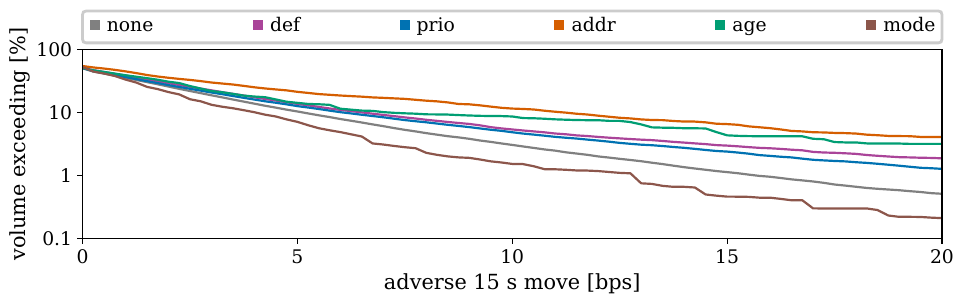}

\caption{For each penalty group, share of its volume followed by a reference-price move exceeding $x$\,bps in the trader's direction over a 15\,s horizon.}
\label{fig:adverse-flow}
\end{figure}

Figure~\ref{fig:adverse-flow}\,a visualizes these curves. At larger thresholds, address- and age-penalized flow exhibit the heaviest informative tails, while mode-penalized flow exhibits the thinnest tail. Reference-price movements exceeding 10\,bps in the trader's direction follow approximately 11.54\%, 8.64\%, and 1.53\% of these groups' volumes, respectively. 

\subsection{Quotes vs. Execution}\label{app:tessera_q_e}

Next, we compare the price a trade would have received if it had executed on the state at the end of the previous block with the price it received in the state it executed on. To that end, for every Tessera WETH/USDC swap we compute the output for the same input twice: the output the contract would have returned in the state at the end of the previous block, i.e., the \emph{quote} an aggregator routing on that state expects, and the output it would have returned in the state immediately before the swap executed, i.e., the \emph{fill} the trade receives.

\begin{table}[t]
\centering\small
\setlength{\tabcolsep}{6pt}
\resizebox{\linewidth}{!}{\begin{tabular}{@{}lrrrr@{}}
\toprule
Cause of difference & quote $>$ fill & fill $>$ quote & identical & volume (\%) \\
\midrule
Nothing changed                  &  0 &  0 & 100 &  37 \\
Different default fee            & 94 &  6 &   0 &  18 \\
Operator updated price, spread, or orderbook config & 39 & 61 &   0 &  37 \\
Earlier swap in the same pool    & 35 &  6 &  59 &   6 \\
Other fee or pricing difference  & 99 &  1 &   0 &   2 \\
\midrule
\textit{All swaps}               & 35 & 24 &  41 & 100 \\
\bottomrule
\end{tabular}}\vspace{2pt}
\caption{Tessera WETH/USDC swaps, comparing the output implied by the state at the end of the previous block with the output implied by the state immediately before execution. Each swap is assigned to the first applicable cause in the order listed, so the causes are disjoint.}
\label{tab:tessera-context-outcomes}
\end{table}

The quote and the fill can differ for several reasons. The operator can post a new price or book, or apply a different spread. Another swap can hit the same pool earlier in the block. The contract can apply a different default fee, and a residual group collects the remaining fee and pricing differences. A swap can be affected by more than one of these, so we assign each to the first applicable cause in the order listed in Table~\ref{tab:tessera-context-outcomes}.

When the applied default fee differs, the trader receives less than quoted on $94\%$ of volume, and the same holds for $99\%$ of volume in the residual group. Operator updates work the other way, with $61\%$ of that volume executing better than quoted. Preceding trades barely matter, since $59\%$ of their volume receives exactly the quoted output. On $37\%$ of volume nothing changes at all and the quote is exact.

\begin{figure}[t]
    \centering
    \includegraphics[width=\linewidth]{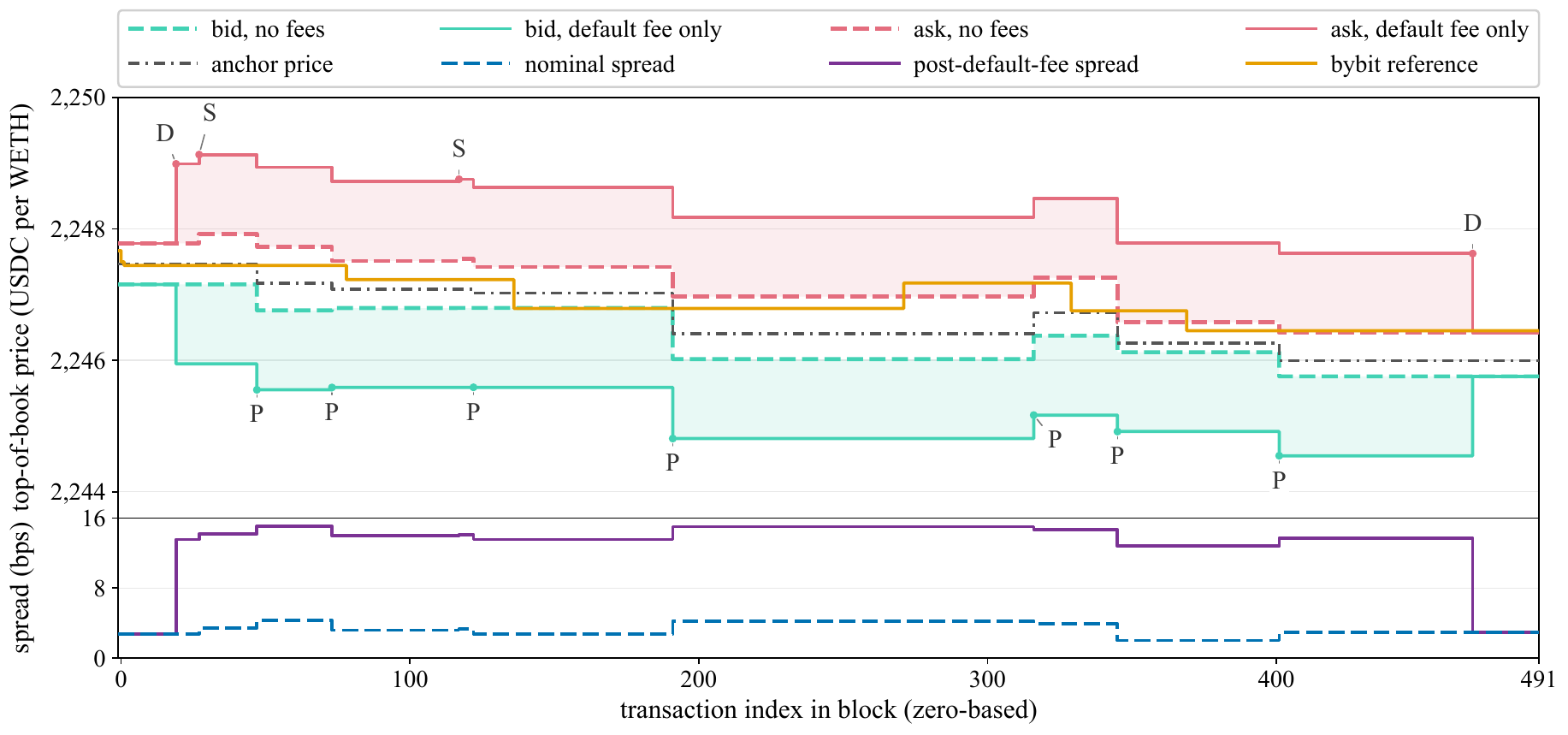}
    \caption{Tessera ETH/USDC prices and spreads within Base block
    50,196,022. Bid and ask prices as well as spreads are shown without fees
    and with only the default fee applied. The anchor price is overlaid.    }
    \label{fig:tessera-intrablock-prices}
\end{figure}

Figure~\ref{fig:tessera-intrablock-prices} illustrates price and default-fee adjustments within an example block and their causes. At the start and the end of the block there is a default fee change marked by $D$. We further show the swaps (marked with $S$) and anchor price and spread updates (marked with $P$). Notice that the default fee is increased before the first swap in this block and decreased after the last swap in the block. This pattern is consistent with propAMM spoofing. The post-default-fee top-of-book spread widened from 2.78 to 13.54 bps when the operator increased the default fee from 0 to 538 ppm at transaction index 19. Near the end of the block, resetting the fee to zero at index 468 narrowed the spread from 13.70 to 2.94 bps.

\begin{figure}[t]\centering
\begin{subfigure}[b]{0.49\linewidth}\centering
  \includegraphics[scale=1]{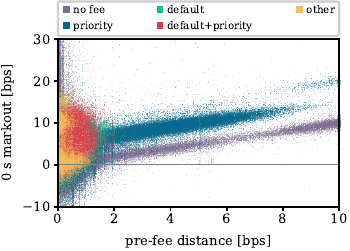}
  \caption{Pre fees}\label{fig:tessera_dist_pre}
\end{subfigure}\hfill%
\begin{subfigure}[b]{0.49\linewidth}\centering
  \includegraphics[scale=1]{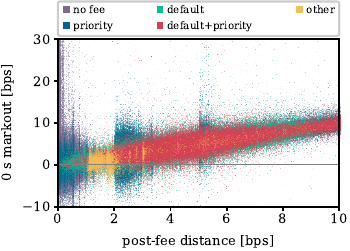}
  \caption{Post fees}\label{fig:tessera_dist_post}
\end{subfigure}
\caption{Tessera's $0$\,s markout against the distance from its own top of book on Base, before and after fees, by which \emph{combination} of fee components the trade pays; the categories are mutually exclusive and the distance is one-sided. The four named combinations cover $99.1\%$ of trades and the remaining fourteen are pooled into \emph{other}. Counting each component wherever it appears, the default fee is charged on $22.7\%$ of trades and the priority fee on $42.7\%$, while $42.1\%$ pay nothing.}
\label{fig:tessera_distance_penalties}
\end{figure}
\section{Tessera vs. Elfomo vs. Metric}\label{app:base_prop_amm}

Next, we compare the pricing mechanisms of Tessera, Elfomo, and Metric for the ETH/USDC pair on Base. The implementations we base our analysis on are \href{https://basescan.org/address/0x6d9dd143e42b6338f4f6a7c0c26d124658f641cb}{\texttt{0x6d9...1cb}} for Tessera, \href{https://basescan.org/address/0x611014f2cccb9d0ddb669dc0b1fa9cd005da647e}{\texttt{0x611...47e}} for Elfomo, and \href{https://basescan.org/address/0x3f63526c786d0f218483788e619a51ba4191b289}{\texttt{0x3f6...289}} for Metric. 

\parhead{Tessera.} Tessera represents liquidity through two directional pricing ladders, akin to traditional orderbooks (Appendix~\ref{app:tessera}). External updates adjust the parameters of the ladders, while swaps change accessible liquidity through directional flow accumulators. Execution first traverses the available ladder liquidity and then applies the combined execution penalty to the resulting output.

\begin{figure}[t]
  \centering
  \captionsetup[subfigure]{font=small,labelfont=bf,skip=1pt,justification=centering}
  \begin{subfigure}[b]{0.32\linewidth}
    \centering
    \includegraphics[width=\linewidth]{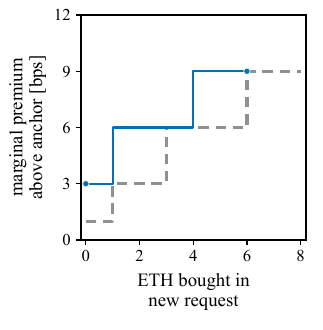}
    \caption{Tessera}
  \end{subfigure}\hfill
  \begin{subfigure}[b]{0.32\linewidth}
    \centering
    \includegraphics[width=\linewidth]{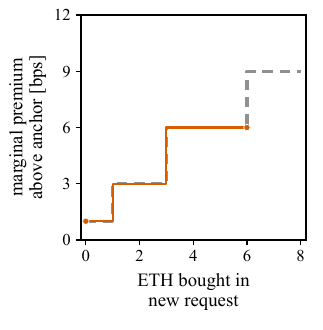}
    \caption{Elfomo}
  \end{subfigure}\hfill
  \begin{subfigure}[b]{0.32\linewidth}
    \centering
    \includegraphics[width=\linewidth]{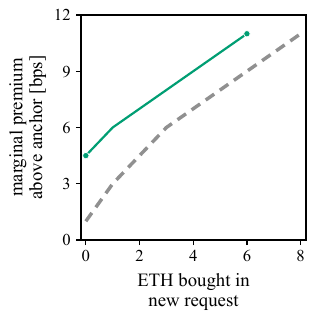}
    \caption{Metric}
  \end{subfigure}
  \caption{Illustrative marginal purchase curves. Dashed lines show the initial
    state and colored lines show a new request after a 2 ETH purchase, with fixed
    reference prices and parameters. Fees and Elfomo's terminal level are omitted.}
  \label{fig:base_pricing_comparison}
\end{figure}

\parhead{Elfomo.} Elfomo also traverses a ladder of price levels, but reconstructs their available capacities depending on current vault balances as well. Its reference combines posted price updates with the raw price stored in WooFi's Wooracle contracts~\cite{woofi_onchain_price_feeds}, with freshness and deviation checks.\footnote{For example, the trace of \href{https://basescan.org/tx/0x5ac4c8333ff813e0559843c3b4ba2f6ad9676c717dbe69f1a3245ffa7ac50529}{\texttt{0x5ac...529}} shows the pricing module calling Wooracle's \texttt{timestamp()} and \texttt{infos(WETH)} getters to fetch WooFi's posted price.} Level, inventory, quote-age, and caller-dependent penalty adjustments enter the prices during traversal, rather than being deducted from the input or the final output as in Tessera's case. A further penalty based on preceding same-side executions in the same block adjusts the final amount. This is based on trade count rather than Tessera-style amount-based flow netting.

\parhead{Metric.} Metric combines a Chainlink ETHUSD data stream (e.g., \href{https://basescan.org/tx/0xe7112aa6a8baa055429dc18ea59d2bf60076fe54b5400e223b24024cfa2afd85}{\texttt{0xe71...d85}}) with separately posted bid and ask quotes (e.g., \href{https://basescan.org/tx/0x57334fcaf535671ba2db37e3504591f50d098a0250014f1acb478a771b5d295c}{\texttt{0x573...95c}}). The quoter constructs a band around the Chainlink-derived reference and combines it with the separately posted bid and ask quotes by taking the lower bid and higher ask. Liquidity is organized into bins with price intervals: marginal prices vary within a bin, and larger trades traverse multiple bins, similar but not identical to how v3-style AMMs behave. Bin reserves and the position within the liquidity curve persist between swaps. The specific penalties and adjustments enter the traversal calculation before the whole fill is calculated. Thus, both Metric and Elfomo incorporate pricing adjustments during traversal, whereas Tessera applies its additional execution penalties after traversal. Furthermore, Chainlink mainly comes into play for Metric during the updates, whereas WooFi comes into play for Elfomo during swap execution.

\section{Markouts by propAMM}\label{app:markout_prop_base_mon}

\begin{figure}[t]\centering
\begin{subfigure}[b]{0.32\linewidth}\centering
  \includegraphics[scale=1]{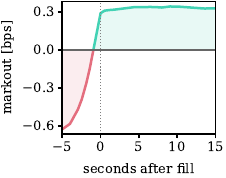}
  \caption{HumidiFi}
\end{subfigure}\hfill
\begin{subfigure}[b]{0.32\linewidth}\centering
  \includegraphics[scale=1]{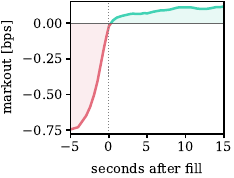}
  \caption{BisonFi}
\end{subfigure}\hfill
\begin{subfigure}[b]{0.32\linewidth}\centering
  \includegraphics[scale=1]{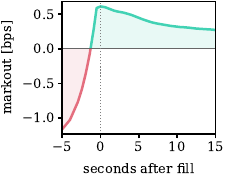}
  \caption{Tessera}
\end{subfigure}
\caption{Gross maker markout of the three largest Solana propAMMs on SOL/USDC against the Bybit microprice $t$ seconds after the fill. Line: notional-weighted mean across days.}
\label{fig:markouts}
\end{figure}

Figure~\ref{fig:markouts} breaks Solana down by venue for the three largest propAMMs. All three are negative $5$\,s before the fill and rise steeply into $t=0$, but they diverge afterwards. Tessera earns the largest markout at $0.56$\,bps $2$\,s after the fill and then gives back part of the gain, falling to $0.28$\,bps by $15$\,s. HumidiFi holds its markout near $0.32$\,bps across the whole horizon. BisonFi earns the smallest at $0.06$\,bps, and is the only one whose markout keeps rising past ten seconds, which likely reflects price prediction that pays off over longer horizons.

\begin{figure}[t]
  \centering
  \begin{minipage}[t]{0.32\linewidth}
    \centering
    \includegraphics[width=\linewidth]{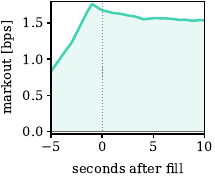}
    \par\smallskip
    {\small\textbf{(a)} Tessera\par}
  \end{minipage}\hfill%
  \begin{minipage}[t]{0.32\linewidth}
    \centering
    \includegraphics[width=\linewidth]{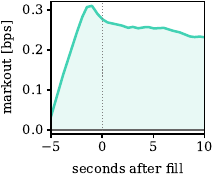}
    \par\smallskip
    {\small\textbf{(b)} Metric\par}
  \end{minipage}\hfill%
  \begin{minipage}[t]{0.32\linewidth}
    \centering
    \includegraphics[width=\linewidth]{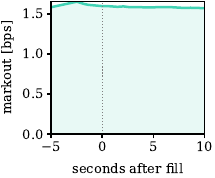}
    \par\smallskip
    {\small\textbf{(c)} Elfomo\par}
  \end{minipage}
    \caption{Volume-weighted ETH/USDC maker markouts against the Bybit ETHUSDT microprice divided by USDCUSDT midpoint.}
  \label{fig:markouts_base}
\end{figure}

Figure~\ref{fig:markouts_base} plots the markouts of the three largest propAMMs on Base. Two differences to the markouts of propAMMs on Solana stand out. First, the markout is positive across the entire window for all three propAMMs, and the rise into $t=0$ is smaller than on Solana, i.e., a move in the reference price triggers a fill less often than on Solana. Elfomo is flat at $1.6$\,bps throughout, while Tessera and Metric rise moderately and reach $1.62$\,bps and $0.26$\,bps 2\,s after the fill. Second, Tessera and Elfomo earn considerably more per fill than any propAMM on Solana, while Metric earns less than all but BisonFi.

\begin{figure}[t]
  \centering
  \begin{minipage}[t]{0.32\linewidth}
    \centering
    \includegraphics[width=\linewidth]{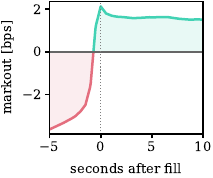}
    \par\smallskip
    {\small\textbf{(a)} Hanji\par}
  \end{minipage}\hspace{10pt}%
  \begin{minipage}[t]{0.32\linewidth}
    \centering
    \includegraphics[width=\linewidth]{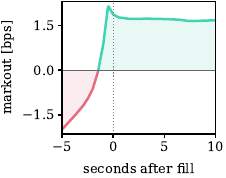}
    \par\smallskip
    {\small\textbf{(b)} Metric\par}
  \end{minipage}
\caption{Volume-weighted MONUSDC maker markouts against the Bybit MONUSDT microprice divided by USDCUSDT midpoint.}
  \label{fig:markouts_monad}
\end{figure}

On Monad, Hanji and Metric behave like the propAMMs on Solana, i.e., their markout is negative ahead of the fill, at $-3.63$\,bps and $-1.99$\,bps 5\,s before, and rises before the fill (Figure~\ref{fig:markouts_monad}). Hanji gives back part of the gain and declines from $2.14$\,bps at the fill to $1.65$\,bps 2\,s later, while Metric declines more modestly from $1.88$ to $1.72$\,bps. 
\section{Markouts by Route Class}\label{app:route_class}
\begin{figure}[t]\centering
\begin{subfigure}[b]{0.32\linewidth}\centering\includegraphics[scale=1]{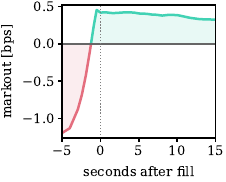}\caption{Solana}\end{subfigure}\hfill
\begin{subfigure}[b]{0.32\linewidth}\centering\includegraphics[scale=1]{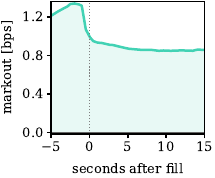}\caption{Base}\end{subfigure}\hfill
\begin{subfigure}[b]{0.32\linewidth}\centering\includegraphics[scale=1]{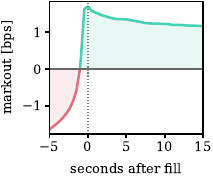}\caption{Monad}\end{subfigure}
\caption{Gross notional-weighted mean maker markout against the Bybit microprice $t$ seconds after the fill, for single fills. We pool all propAMMs per chain, for SOL/USDC on Solana, WETH/USDC on Base and WMON/USDC on Monad.}
\label{fig:markouts_single_venues}
\end{figure}

\begin{figure}[t]\centering
\begin{subfigure}[b]{0.32\linewidth}\centering\includegraphics[scale=1]{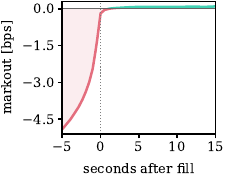}\caption{Solana}\end{subfigure}\hfill
\begin{subfigure}[b]{0.32\linewidth}\centering\includegraphics[scale=1]{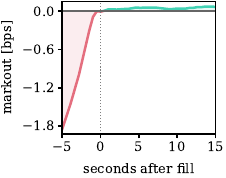}\caption{Base}\end{subfigure}\hfill
\begin{subfigure}[b]{0.32\linewidth}\centering\includegraphics[scale=1]{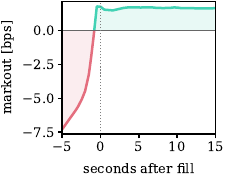}\caption{Monad}\end{subfigure}
\caption{Gross notional-weighted mean maker markout against the Bybit microprice $t$ seconds after the fill, for propAMM fills in propAMM--DEX arbitrage transactions. We pool all propAMMs per chain, for SOL/USDC on Solana, WETH/USDC on Base and WMON/USDC on Monad.}
\label{fig:markouts_prop_dex_venues}
\end{figure}

Figures~\ref{fig:markouts_single_venues} and~\ref{fig:markouts_prop_dex_venues} split the pooled propAMM markouts of Section~\ref{sec:markouts} into the two route classes of Table~\ref{tab:route_compact}. Single fills are profitable on all three chains at $2$\,s, earning $0.41$\,bps on Solana, $0.92$\,bps on Base and $1.44$\,bps on Monad, whereas propAMM--DEX legs earn $0.03$, $0.02$ and $1.55$\,bps. The positive pooled markout is therefore not the product of one transaction type. The classes differ in the pre-fill leg, where propAMM--DEX legs follow a reference-price move of $4.9$\,bps on Solana, $1.9$\,bps on Base and $7.3$\,bps on Monad, against the smaller drift of single fills.

\section{Retail Execution on AMMs}\label{app:retail_AMM}
\begin{figure}[t]\centering
\begin{subfigure}[b]{0.32\linewidth}\centering
  \includegraphics[scale=1]{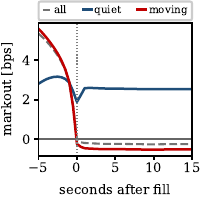}
  \caption{Solana}
\end{subfigure}\hfill%
\begin{subfigure}[b]{0.32\linewidth}\centering
  \includegraphics[scale=1]{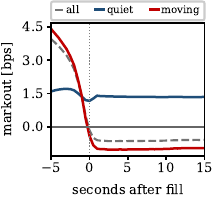}
  \caption{Base}
\end{subfigure}\hfill%
\begin{subfigure}[b]{0.32\linewidth}\centering
  \includegraphics[scale=1]{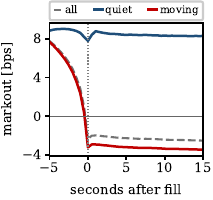}
  \caption{Monad}
\end{subfigure}
\caption{AMM markouts under the same split as Figure~\ref{fig:quiet_prop}. Quiet fills carry only $9\%$ of notional on Solana, $16\%$ on Base and $8\%$ on Monad.}
\label{fig:quiet_AMM}
\end{figure}

Figure~\ref{fig:quiet_AMM} shows that the AMM loss is not uniform across flow. On quiet flow, AMMs earn $2.59$\,bps on Solana, $1.38$\,bps on Base and $8.60$\,bps on Monad at $2$\,s, while on moving flow they lose $0.49$\,bps, $1.00$\,bps and $2.99$\,bps. The pooled figures of Section~\ref{sec:markouts} are averages of the two, dominated by moving flow, which carries $91\%$, $84\%$ and $92\%$ of AMM notional respectively. 

\end{document}